\documentclass[floatfix,aps,prb,twocolumn,groupedaddress,showpacs]{revtex4}

\usepackage{graphicx}

\begin{document}
\title{Acoustic Phonon Anomaly in  MgB$_2$}
\author{R. F. Wood} 
\author{Bo E. Sernelius} 
\altaffiliation[Permanent address: ]{Department of Physics and
Measurement Technology, Link\"oping University, Sweden.}
\author{A. L. Chernyshev} 
\altaffiliation[Also at ]{Institute of Semiconductor
Physics, Novosibirsk, Russia.}

\affiliation{Solid State Division, Oak Ridge National Laboratory,
P.O. Box 2008, Oak Ridge, TN 37831} 

\date{\today}

\begin{abstract}
Recent first principles calculations of the phonon dispersion curves in
MgB$_2$ have suggested the presence of anomalies in some of the curves,
particularly in the longitudinal acoustical (LA) branch 
in the $\Gamma$ to $A$ direction. Similar behavior has been observed
in numerous other superconductors with T$_c$'s higher than those of
standard electron-phonon BCS superconductors.
Phenomenological calculations of the
$\Gamma\rightarrow A$ LA dispersion based on both an acoustical
plasmon and a ``resonant polarization'' mechanism
are given here to emphasize the importance of these similarities.
\end{abstract}
\pacs{63.20.-e, 63.20.Ls, 74.70.Ad, 74.25.Kc, 71.45.Gm}

\maketitle

Single crystals of the new superconductor MgB$_2$ large enough for
inelastic 
neutron scattering experiments have not yet become available, but first
principles calculations of the phonon dispersion curves have appeared.
\cite{Bohnen,Kong}
In Ref.~\onlinecite{Kong}, 
a well-defined anomaly appears in the dispersion of the LA mode
in the $\Gamma$ to $A$ direction.  
This anomalous behavior is remarkably similar to
that found in transition metal\cite{Nakagawa} and transition metal
carbide\cite{Smith} 
superconductors with high transition temperatures.  Indeed, it has often
been considered a signature of high-temperature electron-phonon
superconductivity in these systems.  Here, we call attention to and
emphasize the potential importance of this similarity by carrying out model
calculations in which the anomalies are produced both by an acoustical
plasmon (AP) mechanism and by a parameterized, ${\bf q}$-dependent, interband
contribution to the dielectric response function.

Many superconductors with even moderately high transition temperatures show
some type of anomaly in the phonon dispersion curves when compared to their
low-T$_c$ or nonsuperconducting analogues.  
In this connection, the lattice dynamics in high-T$_c$ cuprate
materials\cite{Pintschovius}  is currently being revisited.
It now appears that there are effects in the optical
modes\cite{McQueeney}  and in the electronic 
structure\cite{Lanzara} that correlate with 
hole doping and that may be related to superconductivity.

\begin{figure}[b]
\includegraphics[angle=270,width=7cm,clip=true]{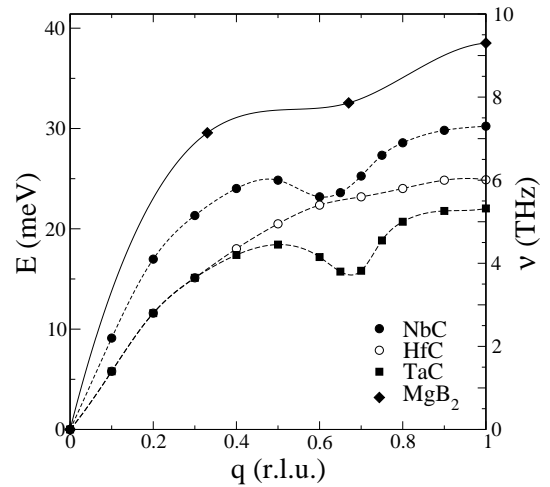}
\caption{Phonon dispersion curves along $(q,0,0)$ for NbC, TaC, and HfC; the
anomalous behavior in the first two is evident.  Data from the recent
first-principles calculations for  MgB$_2$  (Ref.~\onlinecite{Kong}) 
are also shown. Lines are guides to the eye, r.l.u. means reciprocal
lattice units.} 
\label{fig_1}
\end{figure}

We are mainly interested in anomalies in the LA
dispersion curves because they are distinctive and prevalent, occurring in
systems that have no optical modes as well as those that do.  They were
first observed and discussed for Nb (T$_c\sim 10$K) 
more than thirty years ago\cite{Nakagawa}
and somewhat later in the transition metal carbides\cite{Smith}.
Figure \ref{fig_1} shows the LA
 dispersion curves for NbC, TaC, and HfC, which have transition
temperatures of  approximately 11K, 11K, and $<$1K, respectively.  The
anomalies are obvious for the first two materials but do not occur in HfC.
Many other examples of this type of behavior could be given, notably the
elemental materials Nb, V, and Ta where only acoustical modes occur.  Also
shown on Fig. \ref{fig_1} 
are the results from Ref.~\onlinecite{Kong} of recent first-principles
calculations for MgB$_2$ (T$_c=38.5$K).  
While only four points were calculated
in the $\Gamma$ to $A$ direction, they clearly show that the behavior is
anomalous in the sense being used here.

Nakagawa and Woods\cite{Nakagawa} 
found they could reproduce their phonon data for Nb with a Born-von
Karman model utilizing
 numerous force constants. The need for such a complex
model was thought to be indicative of a strong electron-phonon interaction.
Ganguly and Wood\cite{Ganguly} 
observed that the anomalies may be produced by the interaction 
of acoustical plasmons and
phonons.  While it is difficult to reconcile this suggestion with the
electronic band structure of 
Nb, as pointed out by Ruvalds\cite{Ruvalds}, 
the notion that there is a connection
between AP-like 
electronic modes and phonons persisted.  The resonant
electronic polarization arising from the double shell model of Weber
{\it et al.}\cite{Weber}
is somewhat related to the acoustical plasmon mechanism in that an
additional mode which interacts with the ``bare'' phonons is introduced.
First principles calculations\cite{Weber1,Sinha} 
that began to appear subsequently could
replicate the anomalies but the assignments to specific mechanisms 
were often in conflict.  However, the important roles of
electronic polarization and details of the band structure were clear.

The acoustical plasmon concept was introduced by Pines\cite{Pines} 
in 1958 and
subsequently applied to superconductivity by a number of
authors\cite{Radhakrishnan,Gelikman,Frohlich,Rothwarf}. 
It is thought that APs may occur in systems in which heavy and
light mass carriers coexist and the light carriers screen the heavy ones to
produce an acoustical ($\omega\rightarrow 0$ as ${\bf q}\rightarrow
0$) excitation.  Transition metal
systems with the $d$ electrons playing the role of heavy particles and
the $s$ 
electrons the light ones have been long seen as good candidates.
 In other
systems, holes may be the heavy particles and electrons the light ones.  In
the simplest acoustical plasmon treatment of superconductivity in
transition metals, the 
plasmons either replace entirely or simply complement the
phonons in a standard BCS-type theory. An extensive criticism of
 AP-mediated 
superconductivity has been given based on system stability and
other restrictive requirements\cite{Ginsburg,Uspenskii}.
Nevertheless, while to our
knowledge APs
have not been shown to exist in any bulk material (see
also Ref.~\onlinecite{Ruvalds}), the
concept is an intriguing one and based on sound theoretical arguments
regardless of any role it may have in superconductivity. We note that
acoustical plasmons have been proposed\cite{Voelker} 
as existing in MgB$_2$, although not in the context of
the present lattice dynamical discussion and in a much higher frequency
range.

Magnesium diboride has the AlB$_2$ 
structure consisting of hexagonal layers of
Mg interspersed between graphite-like B layers. 
Several calculations\cite{Kortus,Yildirim,Ravindran} establish 
that the occupied bands consist of $\sigma$
bands formed from boron $sp^2$ hybridized orbitals and $\pi$ 
bands formed from the $2p_z$ orbitals directed along the $c$ 
axis.  Interactions with the Mg $3s$
orbitals result in the $\sigma$ bands not being fully occupied, as in 
graphite, which creates holes at the top of the bands.  The various
calculations are in good agreement in showing a weakly dispersing band
in the $\Gamma$ to $A$ direction lying just above the Fermi surface.  
Voelker {\it et al.},\cite{Voelker} 
have provided analytical fits to the bands they calculated, 
with the holes in the $\sigma$ bands having
high effective mass for motion along the $k_z$ direction. This led those
authors to investigate the possible existence of acoustical plasmons.  They
found evidence that they do exist but the validity of their calculation was
subsequently questioned by Ku {\it et al.}\cite{Ku}  
We emphasize that in any case the APs 
considered in these calculations are in a far different frequency range
than those needed here, as will be seen in the following.

We write the ${\bf q}$- and $\omega$-dependent dielectric
function as
\begin{eqnarray}
\label{1}
\varepsilon({\bf q},\omega) = 1 + \alpha^0 + \alpha^l + \alpha^h
+ \alpha^L \ ,
\end{eqnarray}
and look for its zeros in a standard approach.
$\alpha^l$ and $\alpha^h$ are the polarizabilities of 
the two distinct groups of light and heavy carriers and 
$\alpha^0$ contains all other contributions to the
electronic polarizability, e.g., from interband transitions;
$\alpha^L$  is the lattice contribution.
Electronic polarizabilities are calculated in the free electron
approximation:
\begin{eqnarray}
\label{1a}
\alpha^i_{\bf q}(\omega) = \frac{8\pi e^2}{q^2}\ \sum_{\bf k}
\frac{n^i_{{\bf q}+{\bf k}}-n^i_{\bf k}}{\omega+E^i_{\bf k}
-E^i_{{\bf q}+{\bf k}}+i0}\ ,
\end{eqnarray}
where $E^i_{\bf k}$ and $n^i_{\bf k}$ are the free-electron 
energy and Fermi
function for the $i=l,h$ particles, respectively. 

Neglecting $\alpha^L$
for the moment, an approximate expression for the dispersion of the
APs can be obtained in the long-wavelength limit. 
Using the first terms of the expansions given by Lindhard,\cite{Lindhard}
\begin{eqnarray}
\label{2}
\alpha^i_{\bf q}(\omega) \simeq \left\{ \begin{array}{cc}
3\omega_{0i}^2 /q^2v_i^2 \  & \ \ q v_i > \omega\\
-\omega_{0i}^2 /\omega^2  \ & \ \ \ q v_i < \omega \ .
\end{array} \right.
\end{eqnarray}
$v_{i}$ is the Fermi velocity of the $i$th particle, and 
$\omega_{0i}^2= 4\pi e^2n_i /m^*_i$
is the  plasmon frequency, 
with $m^*_i$ the effective mass and $n_i$ the particle density; the
imaginary parts are neglected. 
Substituting these in Eq. (\ref{1}) and defining 
$1+\alpha^0=\varepsilon_0$
gives for the acoustical plasmon frequency, $\omega_{p}(q)$, 
\begin{eqnarray}
\label{5}
\omega_{p}(q)=\frac{v_p q}{\sqrt{1+q^2\lambda^2}}
 \ , \ \ qv_h \ll \omega_p(q) \ll qv_l\ ,
\end{eqnarray}
where $v_p=v_h\sqrt{N_h/3N_l}$ is the acoustical plasmon sound velocity,
$N_i$ is the density of states of the  $i$th particle
at the Fermi level, and  
$\lambda=\sqrt{\frac{\varepsilon_0\hbar^2\pi}{4e^2k_{F,l}m^*_l}}$ is 
 the screening length of light particles. 
The Landau damping of acoustical plasmons due to the continuum of
light particles is neglected.

In a more complete calculation, we use Eqs. (\ref{1}) and 
the full Lindhard expressions for $\alpha^i$.
To include the phonon contribution we use a simple,
approximate expression for the lattice polarizability
$\alpha^L$
\begin{eqnarray}
\label{6}
\alpha^L_{\bf q}(\omega)= -\omega^0_{ph}(q)^2
(\varepsilon^0+\alpha^l_{\bf q}(\omega))/\omega^2 \ ,
\end{eqnarray}
and take in the $\Gamma$ to $A$ direction
\begin{eqnarray}
\label{7}
\omega^0_{ph}(q)=c_0 \sin\left(\frac{\pi}{2}\frac{q}{q_0}\right) \ ,
\end{eqnarray}
where $q/q_0$ is the fractional distance to the BZ 
boundary and  $c_0$ 
is adjusted to give the frequency of the LA phonon in MgB$_2$  
at the zone boundary, as taken from Refs.~\onlinecite{Bohnen,Kong}.
This implies that the ``bare'' phonon frequency
$\omega^0_{ph}$ already contains the screening due to $\varepsilon^0$ 
and the light carriers, hence this screening
is taken out  in Eq. (\ref{6}).

\begin{figure}[t]
\includegraphics[angle=270,width=7cm,clip=true]{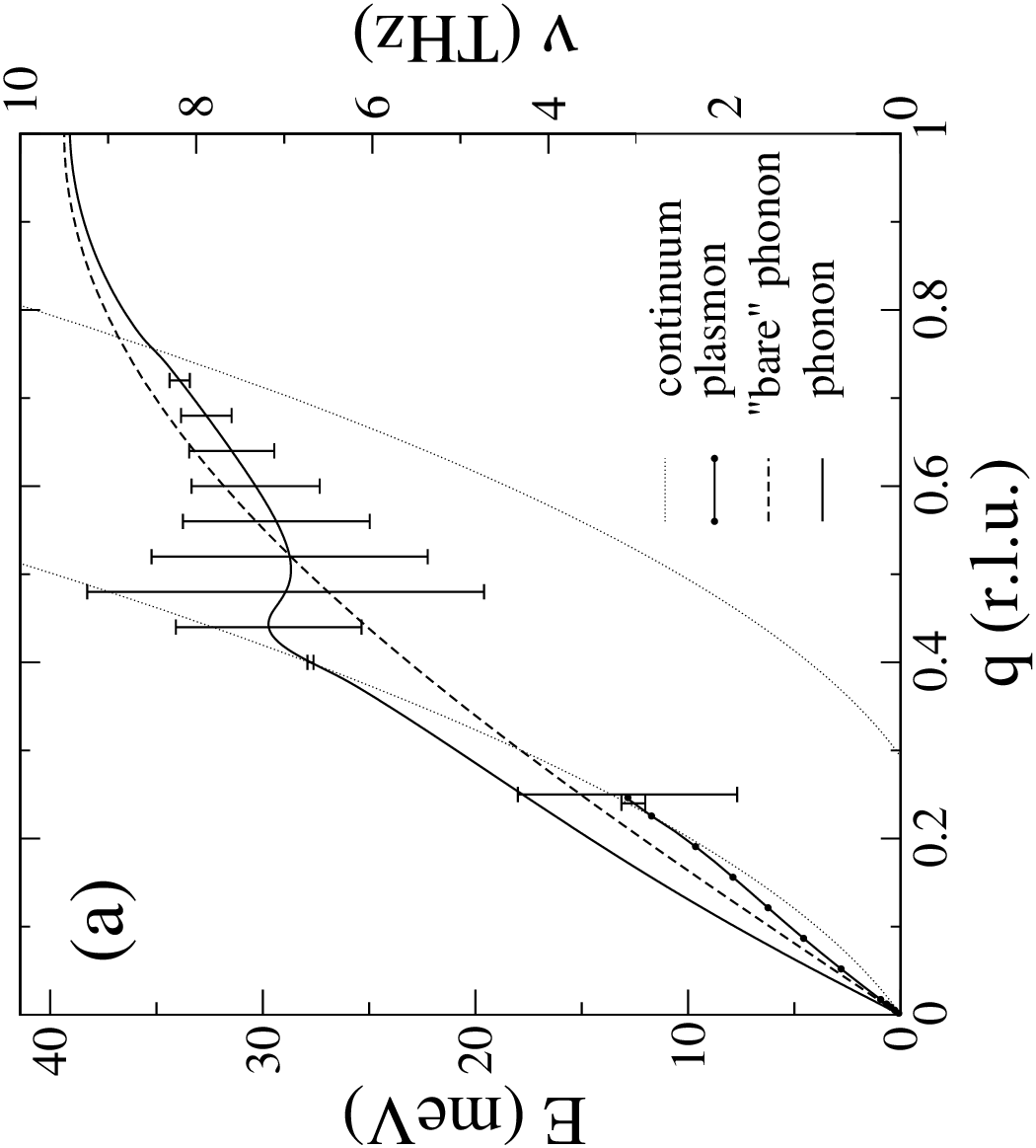}\vskip 0.5cm

\includegraphics[angle=270,width=7cm,clip=true]{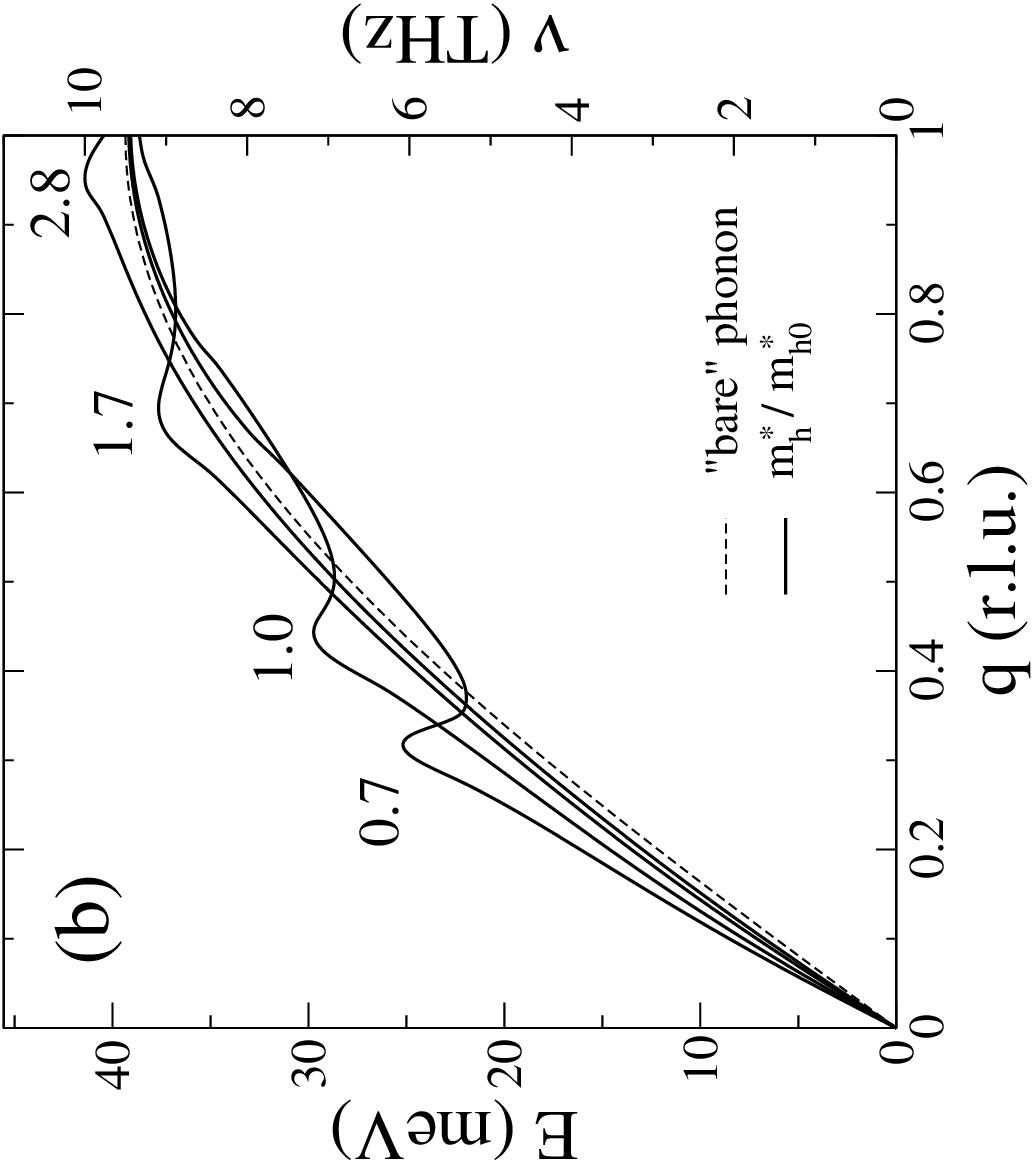}
\caption{(a) An example of the calculation of the anomaly given by the
acoustical plasmon mechanism, as described in the text. (b) The
anomaly as a function of the mass of the heavy particle.} 
\label{fig_2}
\end{figure}

The results of such a calculation are shown in Fig. \ref{fig_2}a.  
The unperturbed or bare phonon
dispersion curve is given by the dashed line and the 
acoustical plasmon by the dotted solid curve.  The latter is
 Landau damped as it enters the heavy particle continuum at $q_c\sim 0.25$.  
Note again that the AP 
mode shown here is calculated {\it without} taking into account the
phonon contribution. Also the bare
phonon is {\it before} the interaction with heavy particles is
included. 
The result of the full calculation
($\varepsilon=\varepsilon_0+\alpha^l+\alpha^h+\alpha^L$) are shown by
the solid line. After the phonon
and heavy particles are allowed to interact, the phonon frequency
rises above the 
unperturbed curve and continues to be affected even after the plasmon is
damped, causing the anomalous behavior to develop.   
Interestingly,
the perturbed phonon remains well-defined until it enters the heavy
particle 
continuum where its width increases rapidly in the region of the anomaly
due to the continuum of heavy particle 
contribution to the imaginary part of $\varepsilon({\bf q},\omega)$.
Neutron scattering experiments should readily pick up this feature if it
exists. We note that this is also the region where the electron-phonon
interaction was found to be strong in 
Ref.~\onlinecite{Kong}. Some of these effects will be discussed in
more detail in a subsequent paper. 

To obtain the curves on Fig. \ref{fig_2}a we took the
mass ratio $m^*_h/m^*_l=20$, and Fermi momentum ratio $k_{F,l}/k_{F,h}=2.4$.
$n_l=10^{21}$ cm$^{-3}$, 
$n_h=0.75\times 10^{20}$ cm$^{-3}$, $v_l=2.4\times 10^7$ cm/s, 
and $v_h=5\times 10^5$ cm/s.  This corresponds to the choice of light mass 
$m^*_l= 1.5 m_0$. Also a value of $\varepsilon_0=30$ was assigned.
As was
the case with Nb, it is difficult to reconcile these parameter values with
the published band structure calculations and this remains a serious
drawback for the application of the acoustical plasmon concept to 
lattice dynamics.

Additional dispersion curves, obtained by changing the effective mass of
the heavy particles over a limited range, are shown on
Fig. \ref{fig_2}b.     By varying
other parameters, a wide variety of behavior can be obtained, which may be
of interest for other materials.

We followed Fr\"ohlich and Rothwarf a bit further by calculating T$_c$ 
from a
modified BCS-like expression,
$kT_c=\hbar\omega_{pm}\exp(-1/F)$. $\omega_{pm}$
was taken to be the
frequency at which the plasmon enters the heavy hole continuum and $F$ was
calculated as described in Refs.~\onlinecite{Frohlich,Rothwarf}. 
Values of T$_c$ quite close to the
experimental value could be found but they were very sensitive to the
relevant parameters which could be changed over a fairly wide range without
greatly changing the dispersion curves. Having in mind the known
restrictions on the acoustical plasmon
mechanism\cite{Ginsburg,Uspenskii}, 
we do not consider
these results of direct, fundamental significance for the high transition
temperatures in MgB$_2$, that is to say, we do not argue that
superconductivity is produced by 
acoustical plasmons. However, 
the results do suggest that the electronic
structure of this material leads to some resonance-like interaction of
the electrons or holes with ``bare'' 
phonons.  How this might come about from another viewpoint is illustrated
by the following calculation.

\begin{figure}[t]
\includegraphics[angle=270,width=7cm,clip=true]{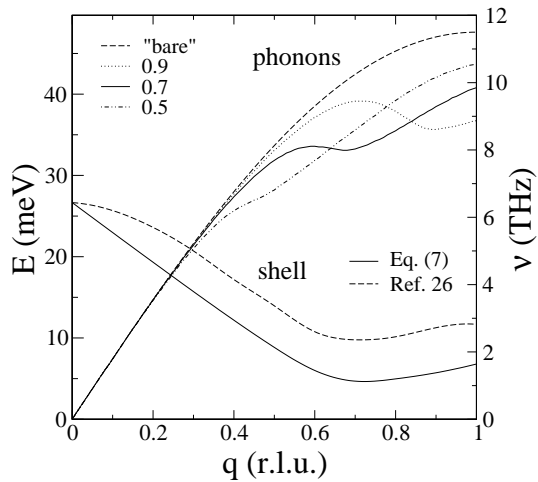}
\caption{Results for the resonant polarization mechanism as described
in the text. The upper curves
are for the phonon and the lower ones are for the supershell frequencies.} 
\label{fig_3}
\end{figure}

  In Eq. (\ref{1}), $\alpha^h$, 
the polarizability of the heavy carriers, which gives rise
to the acoustical plasmons, 
was replaced by a ${\bf q}$-dependent polarization term,
$\alpha^{lh}$, ascribed
here to interband transitions that were previously included in the
constant $\varepsilon^0$.  $\alpha^{lh}(q)$
was chosen to be a ``skewed Lorentzian'' of the form
\begin{eqnarray}
\label{8}
\alpha^{lh}(q)=\frac{a}{1+b^2(q-q_0)^2} \ .
\end{eqnarray}
We initially fixed $q_0$ at $0.7$ 
to locate the maximum in the vicinity of the
dip in the phonon dispersion curve.  Modification of the bare phonon
dispersion and trial and error with the parameters $a$ and $b$ 
readily provided
a satisfactory representation of the anomalous behavior.  The results of
three such calculations for different values of $q_0$ 
are shown by the upper
curves on Fig. \ref{fig_3}.  
We note that the importance of interband transitions has
been emphasized in recent electronic structure calculations and
particularly those of Refs.~\onlinecite{Voelker,Ku}.

An interesting aspect of this calculation is that the $q$-dependence
of $\alpha^{lh}$ correlates
closely with that of the ``supershell frequency'', $\omega_{ss}(q)$, 
in Weber's double-shell model\cite{Weber,Weber1}. 
Smith {\it et al.}\cite{Smith1} calculated $\omega_{ss}(q)$ for TaC by
assigning 
the free electron mass to the supershell. 
The supershell motion was then
retained explicitly in the dynamical matrix rather than being transformed
out as is usually done in the shell model. They found that
$\omega_{ss}(q)$ for the
longitudinal mode had a pronounced,  asymmetrical dip centered at the
position of the anomaly. Loosely speaking, the function of the supershell
is to provide a ``resonant polarization'' mechanism to screen the relevant
part of the dynamical matrix. This implies that the shell frequency should
vary inversely with the square root of this polarization. Calculating the
$q$-dependence of $\omega_{ss}$ in this way from Eq. (\ref{8}) 
and comparing it with the $\omega_{ss}$
extracted from Ref.~\onlinecite{Smith1} 
gives the results shown in the lower part of Fig. \ref{fig_3};
the agreement is obviously quite good. The supershell frequencies have been
normalized to one another at $q=0$ 
and then shifted into the range of phonon
frequencies. Of course, the introduction of a distinct, well-defined mode
for the supershell in this region of $({\bf q},\omega)$  
space is a rather drastic
simplification whose implications are not clear. However, it is remarkable
that the calculations of Ref.~\onlinecite{Ku} 
also point to a sharp
resonance in the vicinity of 2.5~eV  attributed to interband
transitions. In any case, the simple form in Eq. (\ref{8}) does seem
to capture much the same physics contained in the double shell model,
and perhaps also that in less phenomenological treatments.

The calculations of the phonon spectra of MgB$_2$ in
Ref.~\onlinecite{Kong},  which show the anomalies,
used a density functional theory (DFT), full-potential linear response,
LMTO approach.   The
calculations of Ref.~\onlinecite{Bohnen}, 
which show no anomaly, used a mixed
basis pseudopotential method and DFT \cite{elastic_c}.  
While the results of the two
calculations are in general agreement, they differ strongly in the
dispersion of interest here.  Kong {\it et al.}\cite{Kong} 
also  calculated
the electron-phonon interaction and found it to be
particularly large in the vicinity of the anomaly. Since the calculations
of Ref.~\onlinecite{Kong} 
obtained the anomalies, their origin must be embedded in and, in
principle, can be extracted from those calculations.  Conversely, since the
anomalies were not obtained in Ref.~\onlinecite{Bohnen}, 
a comparison of the two approaches
is important from both a computational and a physical standpoint.  As
discussed above, attempts to identify a well-defined mechanism producing
the anomalies in other materials in the 1970s were
inconclusive and MgB$_2$ may now provide an impetus to reconsider their
origins.  It will indeed be interesting to see how well these calculations
fit the experimental lattice dynamical data when crystals large enough
for neutron scattering become 
available.  Of course, any hint of acoustical plasmon effects also would be
of great interest.

To summarize, one recent first-principles calculation of the phonon
dispersion curves in MgB$_2$ shows anomalies in the LA branch similar
to those 
found in other high-T$_c$ electron-phonon 
BCS superconductors; however, another calculation
did not give them.  These anomalies can be calculated using an acoustical
plasmon mechanism but the band parameters involved seem unrealistic.  A
resonant polarization mechanism in which interband transitions give a
${\bf q}$-dependent 
contribution to the dielectric response function is probably
more realistic.  A detailed analysis of the new MgB$_2$ lattice-dynamical
calculations and comparison with earlier calculations for transition metal
materials should be invaluable in giving insight into the underlying
mechanism.

We would like to acknowledge valuable discussions with R.~Fishman,
I.~Mazin, and D.~Singh. This research was supported  by 
ORNL,  managed by UT-Battelle, LLC, for the U.S. DOE under
contract DE-AC05-00OR22725.


\end{document}